\documentclass[12pt]{article}

\usepackage{scicite}
\usepackage[superscript]{cite}

\usepackage{times}

\usepackage{booktabs}
\usepackage{hyperref}
\hypersetup{
colorlinks=true,
breaklinks=true,
urlcolor= RedViolet,
citecolor=blue,
linkcolor=blue,
breaklinks=true
}
\usepackage{breakurl}

\usepackage[labelfont=bf]{caption}
\usepackage{graphicx}
\usepackage[displaymath]{lineno}

\newenvironment{sciabstract}{%
\begin{quote} \bf}
{\end{quote}}

\usepackage[usenames,dvipsnames]{xcolor}

\newcounter{lastnote}

\definecolor{vertdc1}{RGB}{20,89,33}
\definecolor{blood}{RGB}{193,41,41}

\definecolor{vertdc1}{RGB}{20,89,33}
\definecolor{CNRSBlue}{RGB}{26,48,81}
\definecolor{CNRSLightBlue}{RGB}{10,141,167}
\definecolor{DarkRed}{RGB}{212,0,0}
\definecolor{DarkRed2}{RGB}{150,0,0}
\definecolor{Violet}{RGB}{122,47,214}

\title{The floatability of aerosols and waves damping on Titan's seas}

\author
{Daniel Cordier,$^{1}$ Nathalie Carrasco$^{2,3}$\\
\\
\normalsize{$^{1}$Groupe de Spectrom\'{e}trie Mol\'{e}culaire et Atmosph\'{e}rique - UMR CNRS 7331}\\
              \normalsize{Campus Moulin de la Housse - BP 1039,}\\
               \normalsize{Universit\'e de Reims Champagne-Ardenne}\\
               \normalsize{51687 REIMS -- France.}\\[0.3cm]
\normalsize{$^{2}$LATMOS, UMR CNRS 8190, Universit\'{e} Versailles St Quentin,}\\
              \normalsize{UPMC Univ. Paris 06, 11 blvd d'Alembert,}\\
               \normalsize{78280 Guyancourt, France}\\[0.3cm]
\normalsize{$^{3}$Institut Universitaire de France,}\\
              \normalsize{103, bd Saint-Michel,}\\
               \normalsize{75005 Paris, France}\\
}

\usepackage{color,xcolor}

\newcounter{firstbib}

\begin{document} 


\baselineskip24pt


\maketitle


\begin{sciabstract}
%
%
   Titan, Saturn’s largest moon, has a dense atmosphere, together with lakes and 
   seas of liquid hydrocarbons. 
   These liquid bodies, which are in polar regions and up to several hundred kilometres in diameter, 
   generally have smooth surfaces despite evidence of near-surface winds. Photochemically generated 
   organic aerosols form a haze that can settle and potentially interact with the liquid surface. 
   Here we investigate the floatability of these aerosols on Titan’s seas and their potential to 
   dampen waves. We find that the majority of aerosols are denser than the liquid hydrocarbons, 
   but that some could have liquid-repelling properties. From calculation of the capillary forces, we propose 
   that these ‘liquidophobic’ aerosols could float and form a persistent film on Titan’s seas. We 
   numerically model the wave damping efficiency of such a film under the conditions on Titan, 
   demonstrating that even a film one molecule thick may inhibit formation of waves larger than a few 
   centimetres in wavelength. We conclude that the presence of a floating film of aerosols deposited on 
   Titan’s lakes and seas could explain the remarkable smoothness of their surfaces.
\end{sciabstract}

    Titan, the main satellite of Saturn, is the only satellite of the solar system possessing a dense atmosphere. However, the most striking feature 
of Titan is perhaps the presence, of a thick layer of haze, source of inspiration of many works focused on photochemical products and aerosols 
properties{\cite{horst_2017}}. Besides, the presence of oceans of liquid hydrocarbons was conceived in the early eighties, while {\it Cassini} orbiter 
has revealed a collection of seas and lakes in the polar regions of Titan {\cite{stofan_etal_2007}}. These structures involve diameters up to more than 
several hundreds kilometers. Airfall deposits of photochemically produced organics, may coat the surface
of the seas. Potential interactions between haze particles and liquid surfaces are therefore likely. To understand the fate of the aerosols at the surface 
of Titan's lakes, a first parameter to be investigated is their floatability. 
 This issue of flotability is neither anecdotal nor formal. During pre-{\it Cassini} era, in the different context of radar volume
scattering, the near surface properties of Titan's ocean have already been considered{\cite{lunine_1992}}, in the scenario envisaged, particles or 
macromolecules were suspended in the liquid, maintained by some vertical currents.
In the case where some material could
accumulate at the sea surface over time, and build up a layer of some thickness, the presence of a such surface film can
affect the gas, heat and momentum exchange between the sea of the atmosphere. On Earth, it is well known, for centuries, that an oil film damps
the sea surface waves (Aristotle, \textit{Problematica Physica}, 23, no. 38, and historical records mentioned in Ref.~\citenum{alpers_huhnerfuss_1989}).
At the surface of terrestrial oceans, biogenic surface microlayers appear due to  secretion of fish and plankton{\cite{lancelot_mathot_1987}},
and they are detectable by satellite RADAR, through their waves damping effect {\cite{lin_etal_2003}}. 
Therefore, discussing the existence of a possible film, made of atmospheric products, covering some areas of Titan's sea surface, appears
particularly relevant. During flybys other major seas and lakes{\cite{wye_etal_2009,zebker_etal_2014,grima_etal_2017}}, {\it Cassini} RADAR measurements 
have suggested mm-level flatness surfaces.
Infrared ground imagery has also revealed very smooth surfaces for two of them {\cite{stephan_etal_2010,barnes_etal_2011b}}. 
Even if infrared off-specular glints have been observed {\cite{barnes_etal_2014}}, corresponding probably to rougher zones, {\it Cassini} 
observations draw a picture of impressively smooth marine surfaces.
    If the physical and/or chemical properties of some molecules or particles, may lead to 
the appearance of a layer, or microlayer, at the surface of liquid formations, then the properties of 
the seas could be drastically affected. The question touches major properties of this unique case
of ``exo-oceanography''.

\section*{\label{presence}Material that can Sediment to the Surface of Titan}

  To avoid any ambiguity, we reserve the term ``aerosols'' to solid particles that make the thick hazy layer of Titan. 
This photochemical haze 
extends roughly from the surface up to about $1000$ km altitude {\cite{west_etal_2014}} and it is made of aggregates of monomers. Aggregates show
a fractal structure (see Supplementary Fig.~1) and count up to several thousands of monomers {\cite{tomasko_2008b}}; each of them can be 
approximated by a sphere {\cite{curtis_etal_2008}}. The monomers radius determinations, agree for a value around
$50$ nm {\cite{seignovert_etal_2017}}. Each Titan's haze model, based on a microphysical
description, depends on the rate of particle production among several parameters{\cite{rannou_etal_2004}}. 
The aerosols mass production, derived empirically, has its values{\cite{rannou_etal_2004}} spreading around
$\sim 10^{-13}$ kg m$^{-2}$ s$^{-1}$. With the haze layers in steady state, the ``mass production rate'' is also the average ``mass deposit rate'' of aerosols, 
\textit{i.e.} the sedimentation rate over the surface of Titan. The adopted value corresponds, at ground level, 
to one nanometer per year, if we assume a density around $10^{3}$ kg m$^{-3}$. These organic particles should not be surrounded by liquid, in Titan's dry 
regions, while in the most humid ones, aerosols can play the role of nucleation cores for liquid methane droplets formation {\cite{rannou_etal_2006}}.
Even if observational evidences for rainfalls are rare {\cite{turtle_eal_2011a}}; at Titan, the polar regions are recognized to be the wettest from 
climate simulations {\cite{schneider_etal_2012}}. In these regions, the precipitations of liquid methane are governed by the presence 
of small particles (micronic or submicronic) 
known as cloud condensation nuclei, aerosols are very good candidates for this role. In summary,
a certain amount of aerosols should reach the seas as dry particles, whereas the remaining could get the sea embedded in liquid droplets.\\

    Titan's aerosols are the end-products of a complex chemistry, in which a plethora of small molecules is generated.
Some of them are detected by spaceborne instruments {\cite{coustenis_etal_2016}} or by Earth's telescopes {\cite{molter_etal_2016}}.
On the theoretical side, models account for the production of these species {\cite{krasnopolsky_2014}}. Due to local thermodynamic
conditions, some of these compounds can condensate to form either liquid droplets or ice crystals. For instance, the VIMS instrument
aboard {\it Cassini} allowed the detection of micrometre-sized particles of frozen hydrogen cyanide (HCN ice) over Titan’s southern pole
{\cite{dekok_etal_2014}}. The mass flux, to the surface, of these compounds is not negligible. Models indicate a 
mass flux for HCN of the same order of magnitude as that of aerosols{\cite{lavvas_etal_2011c}}.
Hydrogen cyanide is not the only species that can produce organic crystals in the atmosphere, among many others, molecules like
C$_2$H$_2$ and HC$_3$N have also a similar potential{\cite{couturier-tamburelli_etal_2018a}}. 
Nonetheless, HCN appears to be the most abundant. The key point for our purpose,
is the potential propensity of these micron-sized crystals to aggregate with each other into, micron to millimeter-sized
particles, analogs of terrestrial snowflakes {\cite{hobbs_etal_1974}}. The 
physical properties of HCN do not preclude this type of process.
In this perspective, Titan's troposphere could be the scene of ``exotic snowfalls'' 
composed of ``HCN-flakes'' (or C$_2$H$_2$-flakes, ...). Even if CO$_2$ ``snowfalls'' has been considered in the case of 
Mars {\cite{forget_etal_1998}}, perhaps 
curiously this possibility has never be investigated from the point of view of microphysics in the context of Titan.\\
    Finally, two {\it Cassini} instruments detected large molecules
in Titan's thermosphere, with charge/mass
ratios up to{\cite{waite_etal_2007}} $\sim 10,000$. In addition, the presence of polycyclic aromatic hydrocarbons
above an altitude of $\sim 900$ km has been also suggested {\cite{lopez-puertas_etal_2013}}. These facts plead in favor of the 
presence of large molecules at ground level, analogs of terrestrial surfactants{\cite{stevenson_etal_2015a}}.\\
   In summary, we have determined three sources of material that can end at the surface of Titan's hydrocarbon seas:
(1) the haze particles, (2) crystallized organics, (3) large molecules, harboring at least one ``liquidophobic'' function.
%
\section*{Existence and persistence of a floating film}

   Two distinct effects may be invoked when the floatability of an object is questioned: (1) the Archimedes' buoyancy, (2) the
effect of capillary processes.\\


   In an idealized case, the only relevant parameter is the density of monomers material, compared
to the density of sea liquid. In the absence of any wetting effect the liquid penetrates in the whole aerosol free volume. In such
a situation, the fractal structure of aerosols cannot be invoked to introduce an ``effective density'', lower than the monomers one. This is
why we focus on the density of monomers.
   These latter are recognized to be formed by molecules harboring a large number of carbon and nitrogen atoms 
{\cite{gautier_etal_2014}}. As a first guess, we can adopt Earth fossil 
carbon forms, like oil or bitumen, as analogs for monomers the organic matter. Since petroleum industry products are
made of complex mixtures of numerous species, their density is not unique, and for oil{\cite{ancheyta_speight_2007}} ranges
between $0.8$ and $0.95$ g cm$^{-3}$.

   During its descent to Titan's surface, the {\it Huygens} probe made a lot of measurements, and 
the ACP (Aerosol Collector and Pyrolyser)-GCMS (Gas Chromatograph and Mass Spectrometer) experiment analyzed 
the chemical composition of the collected aerosols{\cite{tomasko_2008b}}. 
Their nuclei were found to be made up of N-rich organics, without information about the molecular 
structure. The best estimations of their composition and density are, to date, provided by 
Titan's aerosol laboratory analogues, named ``tholins'' {\cite{sagan_etal_1992}}. 
The few available measurements may be classified into two categories: the high pressure experiments {\cite{horst_tolbert_2013}} producing 
relatively light materials with a density around $\sim 0.8$ g cm$^{-3}$, and the low pressure simulations{\cite{imanaka_etal_2012,brouet_etal_2016}} 
leading to heavier products, with a mean density in the range $1.3 - 1.4$ g cm$^{-3}$. For 
low pressure measurements, individual density determinations can be found down to{\cite{horst_tolbert_2013}} $0.4$ g cm$^{-3}$. 
Concerning ``exotic snows'', densities of solid organics that could
be common at the surface of Titan, may be found in the literature {\cite{cordier_etal_2016b}}: for the most abundant, and less soluble, HCN,
the value should be $\sim 1.03$ g cm$^{-3}$.\\
   Even if the chemical composition of Titan's seas is not known in details, there is a general consensus to consider that
the main components are methane, ethane with some amounts of nitrogen {\cite{cordier_etal_2017a,legall_etal_2016}}. 
In Supplementary Table~1 we have gathered the densities of these species in conditions relevant for Titan's polar surface.
 A quick inspection of this table convinces that, as a general tendency, monomers density should remain above the expected value for the
liquid. In such circumstances, the majority of aerosols particles, or exotic ``snowflakes'', should 
sink to the depths of hydrocarbons seas. This do not exclude the formation of a floating deposit, supported only by Archimedes buoyancy, and formed
by the lightest in the mass spectrum.\\


   We turn now our attention to capillary processes. 
%
%
  It is well known that small bodies heavier than the supporting liquid, including those made of iron, can 
float under the influence of the so-called capillary force. Even some animals, bugs of the family of the \textit{Gerridae} (water striders) 
take advantage of this kind of force to survive at the surface of water {\cite{gao_jiang_2004}}.
   As recalled in Method, the action of a liquid on a tiny object is a function of two parameters: (1) the surface tension $\sigma$ (N m$^{-1}$),
an intrinsic property of the interface liquid-air; (2) the contact angle $\theta_c$, which represents the interaction between the liquid and
the material of the considered object. For $0^{\rm o} \le \theta_c \le 90^{\rm o}$ the liquid is rather ``attracted'' by the solid material, thus called 
``liquidophillic'' in this case. When $90^{\rm o} \le \theta_c \le 180^{\rm o}$, the material is liquid-repellant and is named ``liquidophobic''.
Clearly, an aerosol monomer, seen as a small sphere, may be maintained at the surface solely if the monomer is made of ``liquidophobic'' 
matter.
A simple derivation (see Methods), based on a balance between weight and capillarity forces, leads to the layer thickness
\begin{equation}\label{ethick}
    e \simeq \frac{3 \, \sigma \, |\cos \theta_{\rm c}|}{r \, g_{\rm Tit} \, \rho_{\rm mono}}
\end{equation}
In the case of a perfect non-wetting liquid (\textit{i.e.} for $\theta_{\rm c}= 180^{\rm o}$), a numerical estimate can be obtained for $e$, 
assuming a surface tension $\sigma$ fixed to $2 \times 10^{-2}$ N m$^{-1}$ (see Supplementary Table 2), a radius of monomers at $50$ nm and 
taking $\rho_{\rm mono} = 800$ kg m$^{-3}$, we found $e \simeq 500$ m.
Such an unrealistically large value is the signature of the existence of strong limiting factors. The most obvious limitations are the aerosols
sedimentation rate representing a few nanometer per year, and the idealized poor wettability. This, then, leads naturally to a discussion
concerning expected contact angles.\\ 
      On Titan, maritime surfaces are not the unique place for possible interaction between liquids and solid particles. This kind of interaction
is known to play a crucial role in the formation of cloud particles, which are generated by heterogeneous nucleation{\cite{sanchezlavega}}. 
On the Earth, heterogeneous nucleation on micronic and sub-micronic aerosols
is the dominant mechanism in forming liquid cloud droplets. In the context of Titan, given the large abundance of aerosols, a similar 
microphysics has been proposed for the nucleation of liquid methane droplets or small ethane crystals{\cite{barth_toon_2006}}. 
In these approaches, the contact angle plays a key role that can be easily understood: the more
aerosols are wettable, the more the liquid can spread over its surface and favor the formation of a liquid ``envelope''.
Unfortunately, contact angles are very unconstrained parameters {\cite{rodriguez_etal_2003}},
    what can be found in the literature is either not perfectly relevant{\cite{curtis_etal_2005,curtis_etal_2008}}, or comes from 
informal personal communication{\cite{barth_toon_2006}}.
Except concerning the nucleation of
solid butane, we did not find proper peer reviewed publications providing $\theta_{\rm c}$ values for nucleation onto ``tholins''.
Then, cloud formation models include values close to{\cite{barth_toon_2006}} 
$\theta_{\rm c} \simeq 0^{\rm o}$ and which obviously favors the formation of droplets onto organic aerosols particles.
In other words, microphysics models of clouds assume the existence of ``liquidophilic'' aerosols to play the
role of condensation nuclei, whereas ``liquidophobic'' particles are required to form a floating layer over Titan's lakes surface.\\
  Let us now examine if some clues can be found about the wettability of aerosols.
  The actual chemical composition of Titan's aerosols is not known. Many teams published the global stoichiometry  C$_x$H$_y$N$_z$ of ``tholins'',
which spectral signature is compatible with what is observed at the Saturn
moon. Nonetheless, spectroscopy is not sensitive neither to the detailed chemical composition nor to the exact composition and physical state
of aerosol surface. These surface properties determine the ``liquidophilic'' or ``liquidophobic'' character of aerosols
{\cite{pruppacher_klett_1978}}. 
   During their fall to Titan's ground, particles may also undergo a variety of alterations, due to charging, photolysis or radiolysis
{\cite{courtin_etal_2015,couturier-tamburelli_etal_2018a}}. This 
``aging'', changes the surface properties of
aerosols. This way, their wettability may evolve before they
get to the sea surface. Laboratory measurements show a very low solubility of tholins in non-polar solvents {\cite{carrasco_etal_2009}}. 
An high solubility is generally recognized to be associated with a liquidophilic character. 
Thus, the low solubility of tholins may be regarded as an
indication of a liquidophobia. Similarly, HCN snow may also float due to a strong liquidophobic properties.\\
      Considering the very likely existence of a rich variety of aerosol surface properties, we propose
the presence of both ``liquidophilic'' and ``liquidophobic'' aerosols in Titan's atmosphere. The first family of particles 
sink when they reach the maritime surface, even if they arrive in a ``dry state''. The particles belonging to the second category do not
participate to the clouds formation and float when touching the surface of the sea. 
These kind of particles are good candidates for building up a more or less thick layer at the surface of hydrocarbons seas.
In a sense, the surface of Titan's lakes/seas could retain liquidophobic material.\\
%

     Since the precipitation rates of atmospheric products are small, one might wonder how an organic 
microlayer can build up and be maintained, rather than being destroyed by weathering. Titan surface is a dynamic environment: wind, rain,
fluvial runoff or tides could impede such a formation. It is well known that saltation of particles is much more difficult
from a wet substrate rather than from a dry surface{\cite{lorenz_2014}}. Thus, if lands surrounding seas are wet, the wind should let organic
dusty material lieing down over these terrains, and similarly should not rip marine floating film.
On the contrary, if polar lands are dry, the saltation should be easy and the wind could transport material to sea surface, which could behave 
as a ``wet trap'', leading to an accumulation process. 
Methane rain droplets, or nitrogen bubbles coming from seabed{\cite{cordier_etal_2017a,cordier_ligerbelair_2018}},
may locally disrupt the layer. Thanks to basics physics laws, the momenta associated to the impact of such objects can be estimated 
respectively to $5 \times 10^{-2}$
kg m s$^{-1}$ and $1$ kg m s$^{-1}$, revealing that bubbles could be more efficient than rain droplets. Nevertheless, more specific conclusion
cannot be drawn since the mechanical properties of films are not known. But, bubbles and droplets have a significant difference: droplets
bring to seas material washed up along their fall. Indeed, droplets transport solid particles on which they have nucleated.
If rainfalls are heavy enough, fluvial run-off can also favor the appearance of
surface layers, by transporting material from lands to seas. Finally, according to numerical simulations{\cite{vincent_etal_2016}}, 
Titan's seas undergo a moderate tidal activity. Except along the shores, where material could be periodically deposited and returned to the 
liquid, the tides should not alter any large film due to their large ``wavelength''. Nevertheless, relatively strong tidal currents, through 
the straits may generate some wave fields{\cite{kurata_etal_2016}}.\\
If lakes behave like a trap, an almost continuous shore-to-shore deposit can be expected. In the contrary, where only
a partial coverage is at work, this aspect could introduce some temporal and spatial variabilities in surface properties. 
Even if it seems difficult to destroy floating layers by wind, rain, run-off or tides, these effects could induce migration or 
fragmentation of slicks. Intrinsic properties of floating material could also induce some evolution. Floating 
small objects {\cite{whitesides_boncheva_2002}} can make large structures by self-assembly processes driven by 
lateral capillary interactions.  
Finally, we stress that observations of specular reflections{\cite{soderblom_etal_2012}} over lakes is consistent with a partial and evolving 
film coverage. In the case of a shore-to-shore slick, 
a large range of refractive index values is compatible with glint observations, for which 
the photon fluxes are uncertain due to the lack of knowledge about the optical thickness of the hazy cap.
\captionsetup{labelfont=bf}
%
\begin{figure}[t]
\begin{center}
\includegraphics[angle=0, width=16 cm]{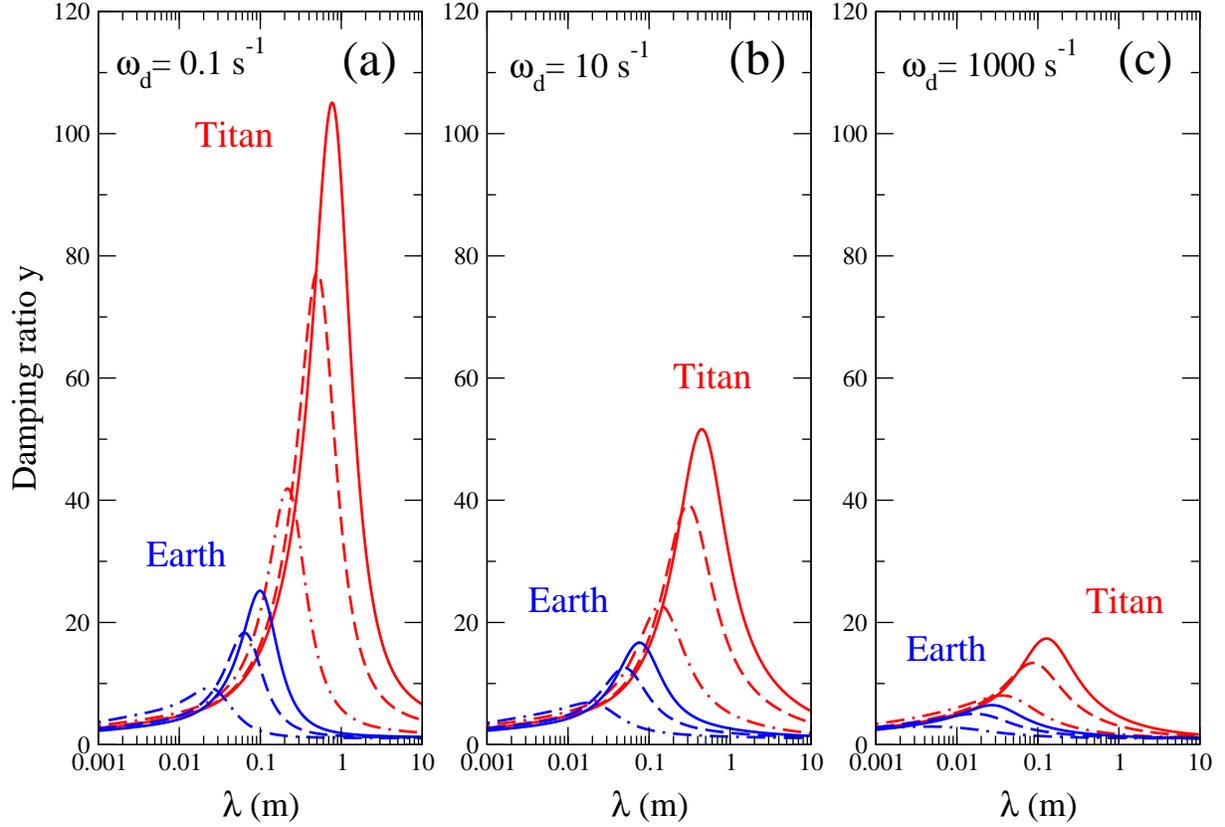}
\caption[]{\label{compaET_y}Comparison of the wave damping efficiency, due to a floating film, between Titan context and under Earth conditions.
           The wave relative damping ratio $y$ caused by a monomolecular film deposited over the 
           surface of a liquid is a function of the wavelength $\lambda$. The three panels
           correspond to different values of $\omega_{\rm d}$ which account for the relaxation time of the
           material forming the slick. In each panel, three values (\textit{i.e.} $36.5$, 
           $21.3$ and $7.3$ in $10^{-3}$ N m$^{-1}$, respectively in solid, dashed and dot-dashed lines) are considered for 
           the coefficient of elasticity in compression $E_{0}$ of the film.
           The parameters concerning the Earth (blue lines) are the gravity $g= 9.81$ m s$^{-2}$, the surface tension $\sigma =
           73$ mN m$^{-1}$, the viscosity $\eta = 10^{-3}$ Pa s, and the density $\rho= 10^{3}$ kg m$^{-3}$, value relevant for 
           liquid water. In the case of
           Titan (red lines), we took $g= 1.352$ m s$^{-2}$ for the gravity, and values expected for liquid methane: 
           $\sigma = 2 \times 10^{-2}$ N m$^{-1}$, $\eta = 2 \times 10^{-4}$ Pa s and $\rho= 452$ kg m$^{-3}$.}
\end{center}
\end{figure}
%

%
\section*{Damping of Sea Waves by Surface Films}
\label{damping}
  The first fully satisfying theoretical explanation has been published in the sixties {\cite{van_den_Tempel_van_de_Riet_1965}}. 
For a monomolecular film, the damping of a wave of initial amplitude $a_0$, after a propagation along a distance $x$ can be written
\begin{equation}
    a(x)= a_0 \, \exp -\Delta \, x
\end{equation}
with $\Delta$ (m$^{-1}$), the damping coefficient, which depends on the wavelength $\lambda$. A ``clean surface'', {\it i.e.} free of slick, has a damping 
coefficient noted $\Delta_0$ (see Method). 
In order to characterize the damping effect of a supernatant film, it is usual to introduce the relative damping ratio
defined as {\cite{alpers_huhnerfuss_1989}}
\begin{equation}
    y(\lambda)=\Delta/\Delta_{0}
\end{equation}
   This ratio depends on the intrinsic mechanical properties of the surface slick, which are represented by $E_0$ (N m$^{-1}$) the modulus
of its coefficient of elasticity, and $\omega_d$ (s$^{-1}$) a parameter accounting for the relaxation time of the layer (see Method for details).
In Fig.~\ref{compaET_y}, we have reported the variations of $y(\lambda)$, employing values representative for monomolecular films 
(\textit{e.g} for hexadecanoic acid methyl ester $E_0= 4.5 \times 10^{-2}$ N m$^{-1}$ and $\omega_d= 22$ rad s$^{-1}$, while for 
oleic acid $E_0= 1.4 \times 10^{-2}$ N m$^{-1}$ and $\omega_d= 38$ rad s$^{-1}$).
Not surprisingly, large viscoelastic modulii $E_0$ produce
a strong damping effect; whereas, long relaxation times (\textit{i.e.} low frequencies $\omega_d$) lead to efficient damping. If we compare the Earth
and Titan, the general tendency is at least a similar damping effect at short wavelengths, and a much stronger effect at 
longer wavelengths. The properties of the sea liquid have also their influence on the waves formation. Except the surface tension $\sigma$ 
(see Supplementary Fig.~2), which has a minor influence, all the other parameters tend to enhance the wave damping at Titan.
Undoubtedly, the sea viscosity $\nu$ has the strongest
effect by, in our example, multiplying the value of $y$ by a factor of $\sim 4$; corresponding to a factor of $\sim \exp 4 \simeq 55$ on the wave
amplitude. 
According to this first approach, Titan seems to be more favorable for a wave damping caused by a monomolecular film, than the Earth, 
because  liquid hydrocarbons have a density and a 
viscosity smaller than that of liquid water.\\
   A monomolecular film is the thinnest blanket that one could imagine, but thicker deposits are also conceivable. A formalism, specifically adapted to
these finite-thickness layers, have also been developed (see Method). In that more general frame, the relative damping ratio $y$ depends explicitly
on the slick thickness $d$, and it firmly increases when $d$ becomes larger. Essentially, results obtained with monomolecular films remain valid 
with thicker ones.\\
   Common observations, and numerous academic studies, show that winds, blowing over water, are found to result in the birth and growth
of waves upon sea surface {\cite{komen_etal_1994}}. The global picture of waves generation can be divided into three physical processes.
First turbulence in the wind produces random stress variations on the surface. These pressure and tangential shear fluctuations give rise
to small wavelets, due to resonances in the wind-sea coupling {\cite{phillips_1957,miles_1957}}. Secondly the waves amplitude is reinforced 
by the air flow, the pressure being maximum on the windward side of the crest and minimum on the leeward side {\cite{miles_1957}}. Finally 
the waves start to interact each with other, exciting longer wavelength modes {\cite{komen_etal_1994}}. Many effects conspire to limit
the wave growth in height and wavelength. For instance, the fetch length over which the wind blows and the so-called ``whitecapping'', affect 
the final spectrum of waves {\cite{komen_etal_1994}}.\\
  It is worth noting that without the generation of the very first ripples, due to air turbulent eddies near the surface, the large waves
cannot be produced, and the surface of the ocean would remain mirror-smooth. An estimation, for the wavelength $\lambda_r$, of these initial 
wavelets caused by resonances is given by {\cite{phillips_1957}}
\begin{equation}
\lambda_r = 2\pi \sqrt{\frac{\sigma}{\rho g}}
\end{equation}
with the notation already adopted in previous paragraphs. In the context of the Earth, this equation leads to $\lambda_r \simeq 1.7$ cm, 
whereas a transposition to Titan yields to a similar value $\lambda_r \simeq 3.4$ cm. Our discussion about the damping rate of waves, 
indicates that a very strong damping could occur at Titan mares, with a maximum efficiency around 
a wavelength of a few centimeters (see Fig.~\ref{influ_d_visc}), depending on the nature and actual properties of 
the floating deposit.
Therefore, if the surface of a Titan sea was covered, at least partially, by such a film/slick;
the onset of wave formation could be impeded, leading to the non-existence of waves at all larger wavelengths, in the corresponding regions.
%
\section*{\label{compatibility}Compatibility of a strong wave damping with observations}
%
     In this paper, we have considered the massive presence of aerosols, an other organic products (large molecules, HCN crystals/snow flakes, ...), 
in the atmosphere of Titan; that sediment to the ground where in polar regions hydrocarbon seas and lakes are observed. The formation of a
more or less thin deposit at the sea surface appears to be plausible. 
   As already mentioned, the off-specular infrared observations may not be in conflict with this scenario: (1) the deposit may be patchy, 
letting free liquid being wavy, and/or (2) the floating layer may itself produce these ``reflections'', if the local deposit has a king of
``roughness'' at infrared wavelengths.
%
%
     Recently, a mechanism has been proposed to explain the occurrence of efficient RADAR reflectors at Ligeia Mare, one of
the main Titan's seas {\cite{hofgartner_etal_2014,hofgartner_etal_2016}}. These, so-called, ``Magic Islands'' could be produced by streams
of nitrogen bubbles rising from the sea depths {\cite{cordier_etal_2017a,cordier_ligerbelair_2018}}. This scenario is not in conflict with
the existence of a thin film at the sea surface: bubbles arriving at the surface could locally break the layer, and, in the same time,
the RADAR-waves transparency of that slick could not prevent the observation of bubbles still in the volume of the sea liquid, as it is
proposed.\\
%
%

  This work has strongly highlighted the need for laboratory studies of interactions between cryogenic liquids, relevant for Titan,
and tholins. Particularly, reliable contact angles determinations are fundamental for the behavior of hydrocarbon seas together with nucleation of
liquid droplets within atmospheric microphysics processes. This new class of experimentations includes studies of surfaces states and
compositions of tholins particles.
  As an extension of preliminary works {\cite{lorenz_etal_2005}}, wind-tunnels may also be used, at room temperature, with liquids and
fine particles or floating films, analog to what is expected on Titan.\\
%
%
%
     Given its, potentially crucial, role in the carbon cycle, floating film/slick could be an important target for possible future 
{\it in situ} explorations{\cite{hartwig_etal_2016}}. And, much more speculatively, it could harbor an original 
``exobiological'' activity.\\

%

\def\sciam{Sci.
  Am.}\def\nature{Nature}\def\nat{Nature}\def\science{Science}\def\natastro{Nat.
  Astron.}\def\natgeo{Nat. Geosci.}\def\natcom{Nat.
  Commun.}\def\pnas{PNAS}\def\AnnderPhys{‎Ann. Phys.
  (Berl.)}\def\icarus{Icarus}\def\pss{Planet. Space Sci.}\def\planss{Planet.
  Space Sci.}\def\ssr{Space Sci. Rev.}\def\solsr{Sol. Syst.
  Res.}\def\expastro{Exp. Astron.}\def\jcis{‎J. Colloid Interface
  Sci.}\def\aap{A\&A}\def\apj{ApJ}\def\apjl{ApJL}\def\apjs{ApJS}\def\aj{AJ}\def\mnras{MNRAS}\def\araa{Annu.
  Rev. Astron. Astrophys.}\def\pasj{Publ. Astron. Soc.
  Jpn.}\def\apss{Astrophys. Space Sci.}\def\pasp{Publ. Astron. Soc.
  Pac.}\def\expastron{Exp. Astron.}\def\asr{Adv. Space
  Res.}\def\astrobiol{Astrobiology}\def\areps{Annu. Rev. Earth Planet.
  Sci.}\def\georl{Geophys. Res. Lett.}\def\jgr{J. Geophys.
  Res.}\def\gca{Geochim. Cosmochim. Ac.}\def\epsl{Earth Planet. Sci.
  Lett.}\def\plasci{Planet. Sci.}\def\ggg{Geochem. Geophys.
  Geosyst.}\def\rmg{Rev. Mineral. Geochem.}\def\tpm{Transport Porous
  Med.}\def\philtrans{Phil. Trans.}\def\faradis{Farad. Discuss.}\def\jcis{‎J.
  Colloid Interface Sci.}\def\jfm{J. Fluid Mech.}\def\physflu{Phys.
  Fluids}\def\pachem{Pure Appl. Chem.}\def\jpcA{J. Phys. Chem.
  A}\def\chemrev{Chem.
  Rev.}\def\nature{Nature}\def\nat{Nature}\def\science{Sci}\def\jced{J. Chem.
  Eng. Data}\def\fpe{Fluid Phase Equilibria}\def\iecr{Ind. Eng. Chem.
  Res.}\def\aichej{AIChE J.}\def\pt{Powder Technol.}\def\etfs{Exp. Therm. Fluid
  Sci.}\def\jgr{J. Geophys. Res.}\def\jcp{J. Chem. Phys.}\def\jcis{‎J.
  Colloid Interface Sci.}\def\jcsft{J. Chem. Soc. Faraday Trans.}

%
\noindent\textbf{\Large Additional information}\\
Supplementary information is available in the on-line version of the paper. Reprints and
permissions information is available on-line at \url{www.nature.com/reprints}. Publisher’s note:
Springer Nature remains neutral with regard to jurisdictional claims in published maps
and institutional affiliations. Correspondence and requests for materials should be
addressed to D.C., e-mail: \texttt{daniel.cordier@univ-reims.fr}\\

\noindent\textbf{\Large Acknowledgements}\\
The authors thanks the anonymous Reviewers who help them to improve the clarity and scientific significance of their work.
N. Carrasco thank the European Research Council for funding via the ERC PrimChem project (grant agreement No. 636829).
This work was also supported by the Programme National de Plan\'{e}tologie (PNP) of CNRS-INSU co-funded by CNES. The authors warmly thank
S\'{e}bastien Lebonnois, Jan Vatant d'Ollone, Tetsuya Tokano and Benjamin Charnay for fruitful scientific discussions.\\

\noindent\textbf{\Large Author contributions}\\
D. Cordier wrote the paper and performed numerical simulations, N. Carrasco provided expertise concerning 
the properties of Titan's aerosols.\\

\noindent\textbf{\Large Competing interests}\\
\noindent The authors declare no competing financial interests.\\

\newpage
\section*{Methods}

\subsection*{Interaction between a monomer and the sea liquid}
\captionsetup{labelfont=bf}

\begin{figure}[!h]
\begin{center}
\includegraphics[angle=0, width=14 cm]{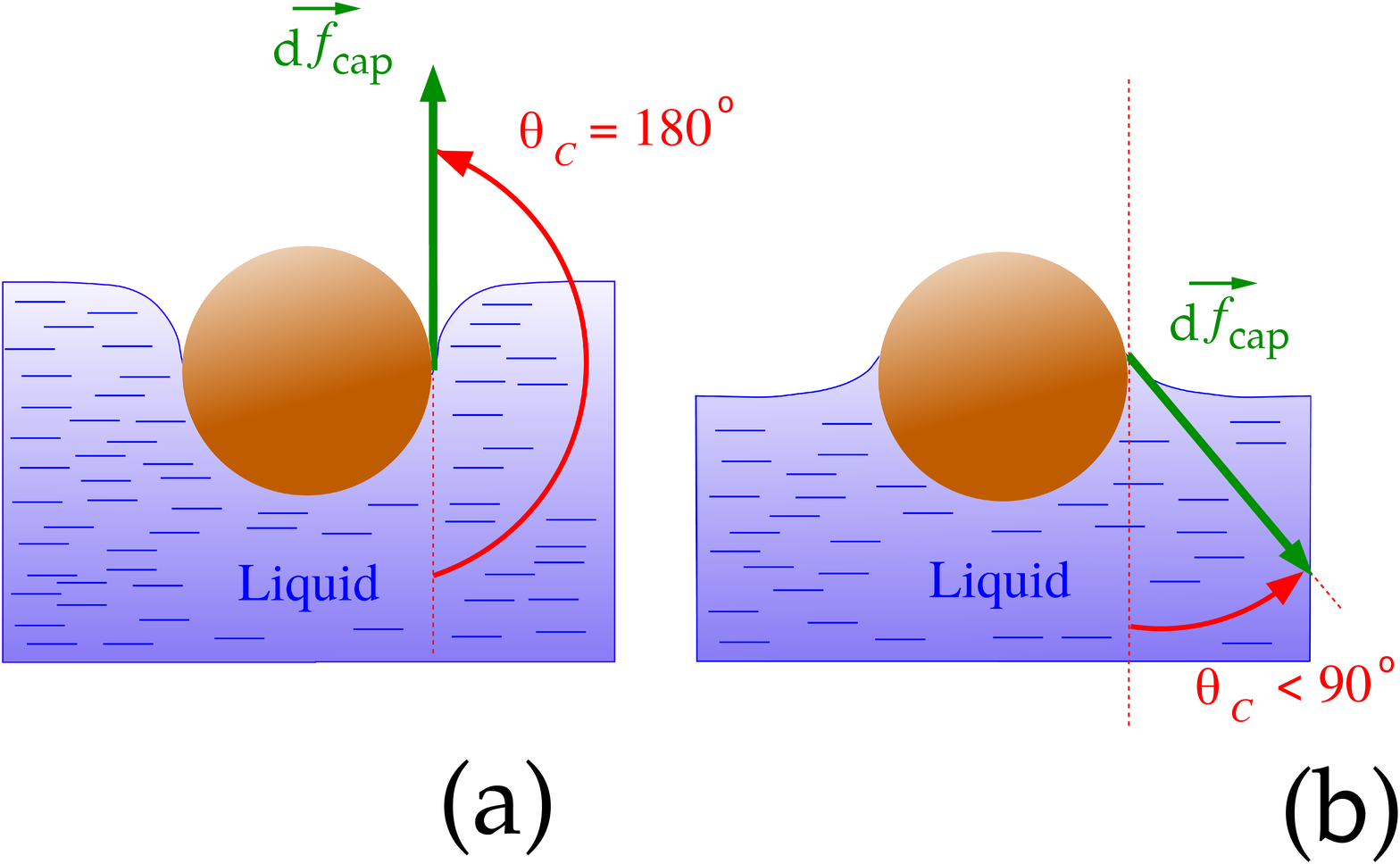}
\caption[]{\label{mono_Angle}A monomer (symbolized by a shaded disk) in contact with a liquid, which is represented in blue. 
          The contact wettability angle between this aerosol monomer and the liquid is noted $\theta_{\rm c}$. 
          (a) The ideal situation of perfect non-wettability of the monomer, in this case the capillarity force sustaining the monomer
          is maximum.
          (b) A wettable monomer: $0^{\rm o} \le \theta_{\rm c} < 90^{\rm o}$, in such a case capillarity cannot produce a floatability effect by
          balancing the gravity.}
\end{center}
\end{figure}
%
  In Fig.~\ref{mono_Angle} we recall some basics about the capillarity forces acting on a tiny object. If the latter is assimilated to
a partially immersed sphere, each elementary portion $\mathrm{d}l$ (m) of the float line undergoes an elementary force, which vertical 
component is given by
\begin{equation}\label{dfcap}
\mathrm{d}f_{\rm cap} = \sigma\cos \theta_{\rm c} \, \mathrm{d}l
\end{equation}
where $\sigma$ is the surface tension of the liquid (N m$^{-1}$), and $\theta_{\rm c}$ is the contact angle. The surface tension is
an intrinsic property of the liquid, it depends on thermodynamic conditions together with the chemical composition of the liquid.
The contact angle accounts for the interactions between the liquid and the solid substrate. The range $0^{\rm o} \le \theta_{\rm c} \le 90^{\rm o}$ corresponds
to high wettabilities: the solid is ``liquidophilic''. The liquid has the tendency to spread out over the entire surface of the solid,
for a  perfect wetting $\theta_{\rm c}= 0^{\rm o}$. A ``liquidophobic'' liquid, \textit{i.e.} presenting a low wettability, leads to
$90^{\rm o} \le \theta_{\rm c} \le 180^{\rm o}$. The perfectly non-wetting case occurs when $\theta_{\rm c}= 180^{\rm o}$. On a flat surface
the liquid tends to form spherical droplets. Figure \ref{mono_Angle} describes the interaction between an idealized spherical monomer and
the liquid phase of a Titan's sea. Clearly, when $0^{\rm o} \le \theta_{\rm c} < 90^{\rm o}$ the monomer is ``attracted'' by the liquid and
consequently sinks into the liquid (see Fig.~\ref{mono_Angle}b). In the contrary, if $90^{\rm o} < \theta_{\rm c} \le 180^{\rm o}$ a vertical
force can balance the effect of gravity, this force is at maximum when the monomer is perfectly non-wettable 
(Fig.~\ref{mono_Angle}a: $\theta_{\rm c}= 180^{\rm o}$). The contact angles are very unconstrained parameters, as discussed in the main text.\\
     For a single monomer, the resulting vertical force $f_{\rm cap}$ (N), produced by capillarity, may approximated by
\begin{equation}\label{fcap}
    f_{\rm cap} \simeq 2\pi r \sigma \cos \theta_{\rm c} 
\end{equation}
where $r$ (m) represents the radius of the monomer. 
This expression assumes that the plane of flotation contains the center of the monomer, which is not necessary the case. Furthermore,
real monomers are built from clusters of large organic molecules and their shape are certainly not perfectly spherical ; therefore
equation (\ref{fcap}) has to be understood as an approximation. Unambiguously, the floatability of the considered monomer is governed by the
value of the angle of contact: if the wettability of the monomer is high (\textit{i.e.} for $0^{\rm o} \le \theta_{\rm c} \le 90^{\rm o}$) the 
capillarity force is pointing down and the monomer sinks. In the opposite, for low wettability (\textit{i.e.} for 
$90^{\rm o} \le \theta_{\rm c} \le 180^{\rm o}$) the capillarity force has a vertical ascending component, that can counterbalance the weight
of the object.


\subsection*{Maximum thickness of a slick supported by capillarity}

  By definition, the first layer of aerosols deposited at the surface of a Titan's sea is in contact with the liquid by
monomers at the lowest positions. We denote $N^{*}$ (N m$^{-2}$) the number, per unit of surface of these monomers located right at the 
surface of the liquid. If $p$ is the ``porosity'' of aerosols, then $N^{*}$ can be estimated by
\begin{equation}
     N^{*} \sim (1-p) \times \frac{1}{s}
\end{equation}
where $s$ (m$^{2}$) is the cross-section of a individual monomer, specifically $s \sim \pi r^{2}$. Thus, the force $F^{*}_{\rm cap}$ (N m$^{-2}$)
per unit of surface, due to the effect of surface tension and acting on the aerosols layer covering the sea, can be derived
\begin{equation}\label{Fetoile}
     F^{*}_{\rm cap} \sim \frac{2 (1-p) \sigma \cos \theta_{\rm c}}{r}
\end{equation}
$M^{*}_{\rm aero}$ (kg m$^{-2}$) is the mass, per unit of surface, of the aerosols deposited at the surface of the liquid.
It can be expressed as
\begin{equation}
     M^{*}_{\rm aero} \sim N^{*} n^{*} \, \frac{4}{3}\pi r^{3} \, \rho_{\rm mono}
\end{equation}
with $n^{*}$ (m$^{-1}$) the mean number of monomers, per unit of length, along the vertical axis. The average density of 
the monomers is noted $\rho_{\rm mono}$ (kg m$^{-3}$). An estimation of $n^{*}$ can be made by adopting $n^{*} \sim e/2r$ with $e$ (m) the total
thickness of the aerosol slick.
The weight, per unit of surface, supported by the liquid, is then $P^{*}_{\rm aero}= M^{*}_{\rm aero} g_{\rm Tit}$ (N m$^{-2}$), where 
$g_{\rm Tit}$ stands for the gravity. With some algebra, we get
\begin{equation}\label{Petoile}
     P^{*}_{\rm aero} \sim \frac{2}{3} \, (1-p) \, e \, \rho_{\rm mono} \, g_{\rm Tit}
\end{equation}
  By equalizing Eq. (\ref{Fetoile}) and Eq. (\ref{Petoile}), we can extract the thickness
\begin{equation}\label{epaisseur}
     e \sim \frac{3 \, \sigma \, |\cos \theta_{\rm c}|}{r \, g_{\rm Tit} \, \rho_{\rm mono}}
\end{equation}
As it could be expected, high surface tension and small monomers favor thick aerosols deposits, while high
gravity together with large densities decrease the value of $e$. Perhaps more surprisingly, this result does not
depend on the ``porosity'' of the aerosols. This behavior is easily explained: both the layer weight, and the surface
capillary force, are proportional to $(1-p)$.

%
\begin{figure}[t]
\begin{center}
\includegraphics[angle=0, width=11 cm]{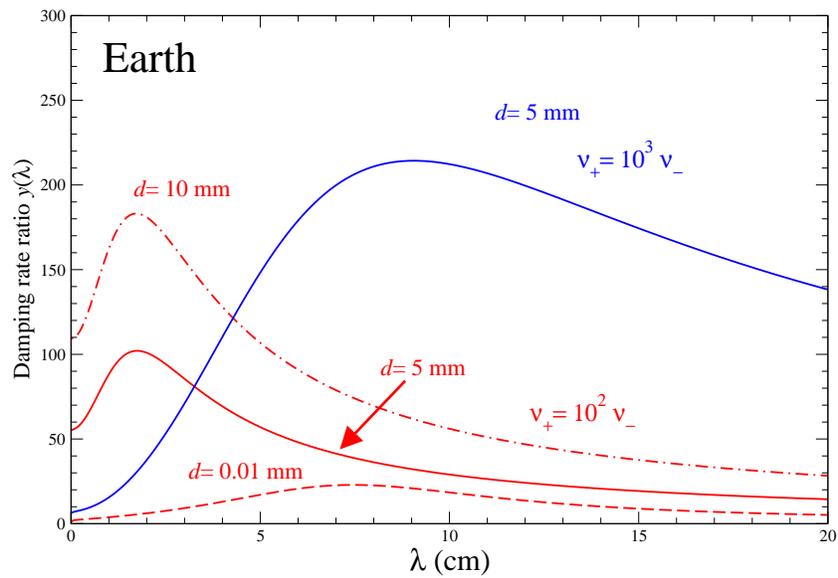}
\caption[]{\label{influ_d_visc}The relative damping ratio $y$ as a function of the wavelength $\lambda$ in the case of a thin finite
           thickness film deposited at the surface of water, \textit{i.e.} in the context of the Earth. The red curves correspond to three
           different thicknesses $d= 0.01$, $5$ and $10$ mm, using a ``standard'' ratio for kinematic viscosities: $\nu_+/\nu_-= 100$. The case
           where $d= 0.01$ mm is approximately similar to the case of a monomolecular floating layer. The solid blue curve shows the influence of
           the film kinematic viscosity $\nu_+$.}
\end{center}
\end{figure}
%
\subsection*{The wave relative damping ratio}

  For a surface, free of floating material, the damping coefficient $\Delta_0$ is given by the Stokes' 
equation{\cite{alpers_huhnerfuss_1989}}
\begin{equation} 
    \Delta_0= \frac{4 k^2 \eta \omega}{\rho g + 3\sigma k^2}
\end{equation}
where $k$ (m$^{-1}$) is the wavenumber related to the wavelength $\lambda$ by the well 
known equation $k= 2\pi/\lambda$. The liquid density, the dynamic viscosity and surface tension are denoted respectively $\rho$ (kg m$^{-3}$), 
$\eta$ (Pa s) and $\sigma$ (N m$^{-1}$). The planet gravity is represented by $g$ (m s$^{-2}$), whereas $\omega$ (s$^{-1}$) is the angular
frequency. In the present context, all these quantities are also constrained by the dispersion relation for gravity-capillary waves
\begin{equation}
    \omega^{2} = g k + \frac{\sigma}{\rho} k^3
\end{equation}
Together with $\Delta$, the actual damping coefficient of the considered sea surface, $\Delta_0$ appears in the relative damping ratio
defined as{\cite{alpers_huhnerfuss_1989}}
\begin{equation}
    y(\lambda)=\Delta/\Delta_{0}
\end{equation}
  The proper mechanical properties of a surface film are a function of its complex viscoelastic modulus $E$ (N m$^{-1}$). 
Its real and imaginary part, respectively 
denoted $E_r$ and $E_i$, as a function of the radial frequency $\omega$ (s$^{-1}$), are given by {\cite{lucassen_giles_1975,cini_1978}}
\begin{equation}
   E_r= E_0 \frac{1 + (\omega_d/2\omega)^{1/2}}{1 + \omega_d/\omega + (2\omega_d/\omega)^{1/2}}
\end{equation}
   and
\begin{equation}
   E_i= E_0 \frac{(\omega_d/2\omega)^{1/2}}{1 + \omega_d/\omega + (2\omega_d/\omega)^{1/2}}
\end{equation}
where $E_0$ is the coefficient of elasticity in compression of the film and $\omega_d$ a parameter which accounts for the relaxation time of the layer.
The relative damping ratio is finally expressed as {\cite{cini_1978}}
\begin{equation}
y= \frac{1 - \frac{\displaystyle E_r-E_i}{\displaystyle 2^{1/2}}a k 
                  + \frac{\displaystyle E_r^2+E_i^2}{\displaystyle 4(2^{1/2})}a^2 k^3 + \frac{E_i}{4} a k^2}{
         1 + (E_r^2+E_i^2) a^2 k^2 -(2^{1/2})(E_r-E_i)a k}
\end{equation}
with $a= k^3/\rho\omega^2$.\\
 The effect of a finite-thickness, {\it i.e.} non-monomolecular, surface deposit has been investigated in the context of heavy fuel slicks and greace ice
{\cite{weber_1987,jenkins_jacobs_1997}}. Up to a thickness of $\sim 10$ $\mu$m, the computed damping rate does not differ essentially
from what we get with zero thickness. In the frame of the formalism, relevant for finite-thickness film, the properties of such a film
is no longer represented by only two parameters. Instead, more specific 
quantities are introduced: the kinematic viscosities of the film $\nu_+$ and that of the bulk liquid $\nu_-$ (m$^2$ s$^{-1}$), 
their respective densities $\rho_+$ and $\rho_-$ (kg m$^{-3}$); together with surface and interfacial tensions. The thickness 
$d$ (m) is also taken into account. The model parametrization is summarized in Supplementary Fig.~3. The behavior of an oil film over water 
at the Earth surface
is illustrated by the damping ratio $y$ for $d=0.01$ mm ({\it i.e.} $10$ $\mu$m) in Fig~\ref{influ_d_visc} .
This curves has to be understood as the reference for what would be produced by a quasi-monomolecular film
{\cite{jenkins_jacobs_1997}}. At all wavelengths, an increase of $d$ leads to a substantially larger relative damping ratio $y$; the conclusions 
of our previous discussion, developed using the simple monomolecular formalism, remain valid and the increase factor of $y$, found in this
frame, must be taken as a minimum value.\\
  Although totally unknown, the intrinsic properties of a possible slick deposited at a Titan's sea surface, should play an important role.
For instance, a high value for the kinematic viscosity $\nu_+$ yields naturally to a much more efficient damping. In fig.~\ref{influ_d_visc}, the
consequences of a ten times more viscous ``oil'' can be seen for a $d= 5$ mm thick layer. As it could be expected, the waves damping appears much
stronger, particularly at long wavelengths.

\noindent\textbf{\Large Code and Data availability} Computer codes and data that support the plots within this paper and
other findings of this study are available from the corresponding author upon reasonable request.\\



%


\end{document}